\documentclass[preprint,11pt,aps,draft,nofootinbib,showpacs,showkeys]{revtex4}
\usepackage{amssymb}

\begin{document}

\title{Maximum information entropy principle and the interpretation of probabilities in statistical mechanics - a short review}

\author{Domagoj Kui\'{c}}
\email{dkuic@pmfst.hr} 

\affiliation{University of Split, Faculty of Science, R. Bo\v{s}kovi\'{c}a 33, 21000 Split, Croatia}

\date{May 27th, 2016}

\begin{abstract}
In this paper an alternative approach to statistical mechanics based on the maximum information entropy principle (MaxEnt) is examined, specifically its close relation with the Gibbs method of ensembles. It is shown that the MaxEnt formalism is the logical extension of the Gibbs formalism of equilibrium statistical mechanics that is entirely independent of the frequentist interpretation of probabilities only as factual (i.e. experimentally verifiable) properties of the real world. Furthermore, we show that, consistently with the law of large numbers, the relative frequencies of the ensemble of systems prepared under identical conditions (i.e. identical constraints) actually correspond to the MaxEnt probabilites in the limit of a large number of systems in the ensemble. This result implies that the probabilities in statistical mechanics can be interpreted, independently of the frequency interpretation, on the basis of the maximum information entropy principle. 
\end{abstract}


\maketitle

\section{Introduction}
From the point of view of predictive statistical mechanics which is based on the maximum information entropy principle, with the exception of quantum mechanical probabilities, there is no reason to consider some particular probability distribution as the true distribution describing the system \cite{jaynes1}. Such a view is in a marked contrast to the interpretation that defines probability only in terms of the limit of a relative frequency of the outcome in an infinite sampling sequence, where the probabilities are therefore factual properties of the observed system \cite{feller}.  From the law of large numbers it follows that the relative frequency of success in a sequence of  e.g. Bernoulli trials (in a sequence of repeated independent trials of an experiment with only two possible outcomes) converges to the theoretical probability. For example, a fair coin toss is a Bernoulli trial where the theoretical probability that the outcome will be heads is equal to $1/2$.
According to the law of large numbers, the proportion of heads in a large number of fair coin tosses will converge to $1/2$ as the number of tosses approaches infinity. This means convergence in probability in the weak form of the law and convergence with probability one in the strong form, where the strong form of the law always implies the weak form of the law \cite{law of large numbers}. Accordingly, the relative frequency is a factual property of the real world that can be measured by repeating a large number of trials, or estimated from the theoretical probability. Probability, on the other hand, is something that we assign to individual events, or we calculate it for the composite events according to the rules (axioms) of probability theory, from the previously assigned probabilities of individual events. 

In different applications of statistical mechanics, we try to predict the results of, or draw inferences from, some experiment that can be repeated indefinitely under what appears to be identical conditions (i.e. on the ensemble of identically prepared systems). Although traditional expositions of statistical mechanics such as \cite{penrose} define the probability as the limiting relative frequency in independent repetitions of a statistical experiment, the relation between frequencies and probabilities, implied by the law of large numbers, in statistical mechanics becomes very complex, because in reality for a macroscopic system, we do not measure the relative frequency of the occurrence of its individual microscopic states in a sequence of infinite or a large number of trials.

In the frequentist interpretation probabilities are always  experimentaly verifiable, and consequently, one of the foundational problems of statistical mechanics would be to derive and to justify the probabilities of microscopic events, in the sense of frequencies in the ensemble of indentically prepared systems, from the first principles i.e. from equations of motion. This is the main problem of ergodic theory approach to statistical mechanics \cite{farquhar,dorfman}. Jaynes presented the opposite view, that if we choose to represent only the degree of our knowledge about the individual system, then there can not be anything physically real in the frequencies in the corresponding ensemble of a large number of systems, nor there is any sense in asking which ensemble is the only correct one \cite{jaynes2}. In the interpretation given by Jaynes, what we call different ensembles corresponds in reality to different degrees of knowledge about the individual system, or about some physical situation. In the argumentation of this viewpoint, Jaynes referred to the statement by Gibbs, according to which the ensembles are chosen only to illustrate the probabilites of events in the real word \cite{jaynes2,gibbs}.

The simplest interpretation of the Gibbs method of ensembles and the MaxEnt formalism follows from the fact that by maximizing the information entropy, which is also known as the uncertainty represented by a probability distribution, subject to given macroscopic constraints, one  predicts just the macroscopic behaviour that can happen  in the greatest number of microscopic realizations (i.e. greatest multiplicity) compatible with those constraints  \cite{jaynes2,jaynes3,jaynes4,jaynes5}. Without going deeper into the problem of interpretation of probabilites, which is even more pronounced in the case of nonequilibrium states, it is more important that the distributions obtained from the application of the principle of maximum information entropy depend  only on the available information and do not depend on arbitrary assumptions related to missing information. If we refer only to predictions, from the same viewpoint one can speak about the objectivity  only in the extent in which the incompleteness of information about the system is taken into account. Consistent with this way of thinking, by applying the principle of maximum information entropy, we come to the relevant statistical distributions, and this is the subject of the paper. 

The structure of the paper is as follows. Section \ref{sec_2} is a brief introduction on the Shannon's concept of information entropy \cite{shannon}, and on the principle of maximum information entropy and MaxEnt formalism formulated by Jaynes \cite{jaynes6,jaynes7}. Section \ref{sec_3} deals with the interpretation of MaxEnt formalism in statistical mechanics as given by Jaynes \cite{jaynes6} and Grandy \cite{grandy,grandy1}. Section \ref{sec_4} introduces the independent interpretation of probabilities in statistical mechanics on the basis of the principle of maximum information entropy. We modify here and extend the analysis given by Jaynes in \cite{jaynes8} and show that it has important consequences for the interpretation of probabilities. Section \ref{sec_5} is the conclusion summarizing the main results of the paper.

\section{Information entropy - measure of uncertainty - and the principle of maximum information entropy} \label{sec_2}

In Shannon's information theory \cite{shannon} the quantity of the form  
\begin{equation}
H(p_1, \dots, p_n) = -K\sum_{i=1}^n p_i\log p_i \ , \label{eq24d}
\end{equation}
has a central role as a measure of information, choice and uncertainty for different probability distributions $p_1, \dots, p_n$. Starting from the understanding that the problem of constructing a communication device depends on the statistical structure of the information that is to be communicated (i.e. on the probabilities $p_1, p_2, \dots, p_n$ of the symbols $A_1, A_2, \dots, A_n$ of some alphabet) Shannon gave until that time the most general definition of the measure of amount of information. Sequences of symbols or "letters" may form the set of "words" of certain length, and the amount of information is measured analogously. Positive constant $K$ in (\ref{eq24d}) depends on the choice of a unit for the amount of information. In real applications expression (\ref{eq24d}), with the logarithmic base $2$ and $K=1$, represents the expected number of bits per symbol necessary to encode the random signal forming a memoryless source. But most importantly, Shannon's interpretation of the function (\ref{eq24d}) is not dependent on the specific context of information theory. He defined the function (\ref{eq24d}) as the measure of {\itshape uncertainty} related to the occurrence of possible events, or more specifically, as a measure of uncertainty {\itshape represented by the probability distribution} $p_1, p_2, \dots, p_n$. This is substantiated by three reasonable properties that are required from such a measure $H(p_1, \dots, p_n)$ that are sufficient to uniquely determine the form of this function: continuity, monotonic increase with number of possibilities in case when all probabilities are equal, and the unique and consistent composition law for the addition of uncertainties when mutually exclusive events are grouped into composite events. Shannon called the function (\ref{eq24d}) the entropy of the set of probabilities $p_1, p_2, \dots, p_n$.

However, we have still not answered an open question on how to determine or to choose the appropriate probability distribution for a particular problem or a system. The principle of maximum information entropy (MaxEnt) was formulated by Jaynes \cite{jaynes6,jaynes7} as a general criterion for construction of the probability distribution when the available information is not sufficient for the unique determination of the distribution. This principle is based on the following rationale: maximization of the information entropy (the uncertainty) subject to given constraints includes in the probability distribution only the information represented by these constraints. Therefore, predictions derived from such a probability distribution depend only on the available information and do not depend on arbitrary assumptions related to missing information.

The mathematical formulation of this principle is known as the MaxEnt algorithm. Let's consider it on the following example. Let the variable $x$ takes $n$ values $\{x_1, \dots, x_n\}$ with probabilities $\{p_1, \dots, p_n\}$ and the only data available are given by the expectation values of the functions $f_k(x)$:
\begin{equation}
F_k = \langle f_k (x)\rangle = \sum_{i = 1}^{n} p_i f_k (x_i) \ , \qquad k = 1, 2, \dots, m < n \ .  \label{eq_expt_Fk_p}
\end{equation}
Probability distribution must also satisfy the normalization condition
\begin{eqnarray}
\sum _{i=1}^n p_i = 1 \ ,  \qquad  \qquad \left (p_i \geq 0, \quad i = 1, 2, \dots, n \right ) \ . \label{eq_cond_norm_nonneg}
\end{eqnarray} 
In most cases the available information given by the set of equations (\ref{eq_expt_Fk_p}) is far less then sufficient for the unique determination of the set of probabilities $\{p_1, \dots, p_n\}$,  i.e. $m << n -1$. In such cases, probability distribution $\{p_1, \dots, p_n\}$ is determined by applying the MaxEnt principle. Probability distribution $\{p_1, \dots, p_n\}$ for which the information entropy (\ref{eq24d}) is maximum subject to the constraints (\ref{eq_expt_Fk_p}) is found by the method of Lagrange multipliers, i.e. by maximizing the function
\begin{eqnarray}
I  & = & - \sum_{i=1}^n p_i \log p_i - (\lambda_0 - 1)\left (\sum _{i=1}^n p_i -1\right ) \nonumber\\
& & - \sum_{k = 1}^m \lambda _k \left (\sum_{i = 1}^{n} p_i f_k (x_i) - F_k\right ) \ ,
\end{eqnarray}
where $\lambda _0 - 1$, $\lambda _k$, $k=1, 2, \dots, m$, are the Lagrange multipliers. In this way we obtain the MaxEnt probability distribution
\begin{equation}
p_i = \frac{1}{Z}\exp \left \{ - \sum_{k = 1}^m \lambda _k f_k (x_i)  \right \} \ , \qquad i = 1,2, \dots, n \ .  \label{eq_maxent_p_i_Z_lk}
\end{equation}
The normalization factor $Z = e^{\lambda _0}$ which is also known as the partition function is given by
\begin{equation}
Z \equiv  Z(\lambda _1 , \dots , \lambda _m) = \sum_{i = 1}^{n}\exp \left \{ - \sum_{k = 1}^m \lambda _k f_k (x_i)  \right \} \ .   \label{eq_maxent_part_fun_l_k}
\end{equation}
The expectation values of the functions $\langle f_k (x)\rangle = F_k $, $k = 1,2, \dots, m$, given by the conditions (\ref{eq_expt_Fk_p}), are equivalently given also by
\begin{equation}
F_k = \langle f_k (x)\rangle = - {\partial \log Z(\lambda _1 , \dots , \lambda _m) \over \partial \lambda _k} \ , \quad k = 1,2, \dots, m \ . \label{eq_F_k_expct_f_k_dlogZ_dl_k}
\end{equation}
Let's assume that set of $m + 1$ equations consisting of $m$ equations (\ref{eq_expt_Fk_p}) and the equation (\ref{eq_cond_norm_nonneg}) is consistent and that these equations are linearly independent. Then, using (\ref{eq_maxent_p_i_Z_lk}) and solving this set of equations, one can determine the Lagrange multipliers $\lambda _k$, $k=1,2, \dots, m$, as single-valued functions $\lambda _k(F)$ of the expected values $F = (F_1, \dots , F_m) $. The proof is given in \cite{kuic}. Then, by introducing the MaxEnt probability distribution (\ref{eq_maxent_p_i_Z_lk}) in the expression (\ref{eq24d}) for information entropy, the maximum of information entropy subject to the conditions (\ref{eq_expt_Fk_p}) and (\ref{eq_cond_norm_nonneg}) is obtained as the function of the expected values $F = (F_1, \dots , F_m) $:
\begin{equation}
(S_I)_{\mathrm{max}} =  \log Z(\lambda _1 , \dots , \lambda _m)  + \sum_{k = 1}^m \lambda _k  F_k = S(F_1 , \dots, F_m) \ . \label{eq_S_I_max_eq_S_F}
\end{equation}
Assuming that the functions $\lambda _k(F)$, $k=1,2, \dots, m$, are continuously differentiable  (or at least piecewise smooth),  from (\ref{eq_F_k_expct_f_k_dlogZ_dl_k}) and (\ref{eq_S_I_max_eq_S_F}) it follows that  
\begin{equation}
\lambda _k = {\partial S (F_1 , \dots, F_m) \over \partial F_k }\ , \qquad  k = 1,2, \dots, m \ .  \label{eq_L_eq_dS_F_dF}
\end{equation}
From equations (\ref{eq_F_k_expct_f_k_dlogZ_dl_k}), (\ref{eq_S_I_max_eq_S_F}) and (\ref{eq_L_eq_dS_F_dF}) it is obvious that the functions $\log Z(\lambda _1 , \dots, \lambda _m)$ and  $S(F_1 , \dots, F_m)$ are mutually related by a Legendre transformation. Functions related in this way contain the same information but it is expressed through different variables.

Furthermore, functions $\log Z(\lambda _1 , \dots, \lambda _m)$ and $S(F_1 , \dots, F_m)$ give, in a simple way, the variances and covariances of the functions $f_k(x)$, $k = 1,2, \dots, m$. Using (\ref{eq_maxent_p_i_Z_lk}), (\ref{eq_maxent_part_fun_l_k}) and (\ref{eq_F_k_expct_f_k_dlogZ_dl_k}) one obtains
\begin{eqnarray}
{\partial^2 \log Z(\lambda _1 , \dots , \lambda _m) \over \partial \lambda_l \partial \lambda_k } & = & -{\partial F_k \over \partial  \lambda_l } = -{\partial F_l \over \partial  \lambda_k } \nonumber\\
& = & \langle f_k (x) f_l(x) \rangle - \langle f_k (x)\rangle \langle f_l(x) \rangle \nonumber\\ 
& = & - A_{kl} , \quad k,l =1,2, \dots, m , \label{eq_dlogZ_dL_ldL_k_eq_A_kl}
\end{eqnarray}
where $A$ is a symmetric matrix, $A_{kl} = A_{lk}$. In a similar way, using (\ref{eq_L_eq_dS_F_dF}) one obtains
\begin{eqnarray}
{\partial^2 S(F_1 , \dots, F_m)\over \partial F_l \partial F_k} & = & {\partial \lambda_k  \over \partial F_l} = {\partial \lambda_l  \over \partial F_k} \nonumber\\
& = & B_{kl}\ , \qquad k,l =1,2, \dots, m \ , \label{eq_dS_F_dF_k_dF_l_eq_B_lk}
\end{eqnarray}
where $B$ is also a symmetric matrix, $B_{kl} = B_{lk}$. Then, from (\ref{eq_dlogZ_dL_ldL_k_eq_A_kl}), (\ref{eq_dS_F_dF_k_dF_l_eq_B_lk}) and the chain rule for derivatives, it follows that
\begin{equation}
{\partial \lambda _j \over \partial \lambda _l} = \sum_{k=1}^m {\partial \lambda _j \over \partial F_k}{\partial F_k  \over \partial \lambda _l } = B_{jk} A_{kl} = \delta _{jl}\ , \quad j,l =1,2, \dots, m \ , 
\end{equation}
and similarly,
\begin{equation}
{\partial F _j \over \partial F _l} = \sum_{k=1}^m {\partial F _j \over \partial \lambda _k}{\partial \lambda _k  \over \partial F_l } = A_{jk} B_{kl} = \delta _{jl}\ , \quad j,l =1,2, \dots, m \ . 
\end{equation}
Therefore, the matrices given by (\ref{eq_dlogZ_dL_ldL_k_eq_A_kl}) and (\ref{eq_dS_F_dF_k_dF_l_eq_B_lk}) are inverses, $A^{-1} = B$.

Elements of the matrix $A$ are the second partial derivatives of the functions $\log Z(\lambda _1 , \dots , \lambda _m)$ and represent the measure of the expected dispersion and mutual correlation of the functions $f_k(x)$, $k = 1,2, \dots, m$. Diagonal elements of the matrix $A$ give as the notion about the deviation  of the variables $f_k(x)$ from their expectation values $\langle f_k (x) \rangle$. Furthermore, from (\ref{eq_maxent_p_i_Z_lk}), (\ref{eq_maxent_part_fun_l_k}) and (\ref{eq_F_k_expct_f_k_dlogZ_dl_k}), it follows that the covariance of some other function $g(x)$ with the function $f_k(x)$ is obtained as
\begin{equation}
- {\partial \langle g(x) \rangle \over \partial \lambda _k} = \langle g(x)f_k(x) \rangle - \langle g(x)\rangle \langle f_k(x) \rangle , \ \ \ k = 1,2, \dots, m .
\end{equation}

\section{Interpretation of MaxEnt formalism in statistical mechanics} \label{sec_3}

It clear that the MaxEnt probability distribution (\ref{eq_maxent_p_i_Z_lk}) has the same form as Gibbs ensemble probability distributions from equilibrium statistical mechanics. This is not surprising since the rationale of the Gibbs method of constructing ensembles was to assign that probability distribution which, while agreeing with what is known (i.e. the data given by constraints), gives the least value of the average index (logarithm) of probability of phase i.e. $\sum_{i=1}^n p_i\log p_i$ \cite{jaynes2,gibbs}. This procedure has lead Gibbs to the canonical ensemble for closed systems in thermal equilibrium with the environment, the grand canonical ensemble for open systems, and an ensemble for a system rotating at a fixed angular velocity. However, MaxEnt formalism represents a general method of statistical inference which is applicable, in principle, to all problems where only incomplete and partial information about the problem is available. Equations from the last section represent the generic form of the MaxEnt formalism. To give them a physical interpretation they should be put in the context of some specific physical situation. Since the Lagrange multipliers $\lambda = (\lambda _1 , \dots, \lambda _m)$, under certain conditions, are single-valued functions of the expected values $F = (F_1 , \dots, F_m)$, and at the same time the only parameters in the MaxEnt probability distribution, physical interpretation of these quantities is of special pertinence in that sense. 

It will now be shown that the physical interpretation of the Lagrange multipliers follows from the relation describing the changes of the expected values. Values of the functions $f_k(x_i)$, $k = 1,2, \dots, m$, associated with the values  $x_i$ of the variable $x$, $i = 1,2, \dots, n$, can represent the eigenvalues of some specific physical quantities, for example energy eigenvalues $E_i$, or eigenvalues of the quantities from the set of compatible quantities. Let us assume that the small change in the expectation values $\langle f_k(x) \rangle$ is done by the small change of the functions $f_k(x_i)$ and the probabilities $p_i$, 
\begin{equation}
\delta \langle f_k(x) \rangle = \sum_{i=1}^{n} p_i \delta f_k(x_i) +  \sum_{i=1}^{n} f_k(x_i)\delta p_i \ , \ \ k = 1,2, \dots, m \ . \label{eq_d_expt_f_k_eq_var}
\end{equation}
Here, $\delta \langle f_k(x) \rangle$ is the change of the expectation value $\langle f_k(x) \rangle$ and $\langle \delta f_k(x)\rangle = \sum_i p_i \delta f_k(x_i) $ is the expectation value of the change of $f_k(x) $. Their difference depends on the changes in the probabilities $\delta p_i$, 
\begin{equation}
\delta \langle f_k(x) \rangle - \langle \delta f_k(x)\rangle = \sum_{i=1}^{n} f_k(x_i)\delta p_i \ , \qquad k = 1,2, \dots, m \ . \label{eq_df_k_expt_expt_df_k_ind_term}
\end{equation}
The change of information entropy $S_I$ is equal to
\begin{equation}
\delta S_I = - \sum_{i=1}^{n} \delta p_i \log p_i \ . \label{eq_dS_I_eq_var}
\end{equation}
Introducing the MaxEnt probabilities (\ref{eq_maxent_p_i_Z_lk}) for $\{p_i\}$ in (\ref{eq_dS_I_eq_var}) and using (\ref{eq_cond_norm_nonneg}) and (\ref{eq_df_k_expt_expt_df_k_ind_term}), one obtains
\begin{eqnarray}
\delta  S & = & \sum_{k = 1} ^m \sum_{i=1}^{n} \lambda _k f_k(x_i)\delta p_i \nonumber\\
& = & \sum_{k = 1} ^m \lambda _k \left (\delta \langle f_k(x) \rangle - \langle \delta f_k(x)\rangle\right ) \ . \label{eq_dS_max_eq_dp_i_f_k}
\end{eqnarray}
Assuming that $\{p_i + \delta p_i\}$ is also a MaxEnt probability distribution, equation (\ref{eq_dS_max_eq_dp_i_f_k}) then gives the change of the maximum of information entropy due to the change in expected values (i.e. the constraints). The meaning of (\ref{eq_dS_max_eq_dp_i_f_k}) is simple to understand, if we introduce
\begin{eqnarray}
&& \delta Q_k = \sum_{i=1}^{n} f_k(x_i)\delta p_i = \delta \langle f_k(x) \rangle - \langle \delta f_k(x)\rangle \ , \cr\nonumber\\ 
&& k = 1,2, \dots, m \ , \label{eq_Q_k_d_expt_f_k_expt_d_f_k} 
\end{eqnarray}
and then using this write $\delta  S$ in the form
\begin{equation}
\delta  S = \sum_{k = 1} ^m \lambda _k \delta Q_k \ . \label{eq_dS_sum_k_L_k_Q_k}
\end{equation} 
Equation (\ref{eq_Q_k_d_expt_f_k_expt_d_f_k}) suggests the interpretation that was given by Jaynes \cite{jaynes6} and Grandy \cite{grandy,grandy1}. The expectation value $\langle \delta f_k(x)\rangle $ of the change $\delta f_k(x)$ is the corresponding generalized work. The remaining part of the change $\delta \langle f_k(x) \rangle$ of the expectation value $\langle f_k(x) \rangle$ comes from the change in the probability distribution $\{p_i\}$ and represents the generalized heat $\delta Q_k$ for the quantity $f_k(x)$. If the function $f_k(x)$ is such that $f_k(x_i) = E_i$ for all  $i$, then $\delta Q_k$ is the heat in the usual sense. Grandy \cite{grandy,grandy1} interpreted the equations (\ref{eq_Q_k_d_expt_f_k_expt_d_f_k}) as the general rule in the probability theory, whose special case is the first law of thermodynamics. Indeed, for a macroscopic system, if $f_k(x_i) = E_i$ for all  $i$, then the corresponding equation (\ref{eq_Q_k_d_expt_f_k_expt_d_f_k}) has the form of the first law of thermodynamics
\begin{equation}
\delta Q = \delta \langle E \rangle - \langle \delta E \rangle = \delta U - \delta W \ , \label{eq_F_L_T_dQ_eq_dU_dW}
\end{equation}  
where $\langle E \rangle = U$ is the internal energy of the system and $\delta W = \langle \delta E \rangle$ is the work done on the system. According to \cite{grandy,grandy1}, the heat $\delta Q$ is the energy transfered through the degrees of freedom over which we don't have control, while the work $\delta W$ is the energy transfered through the degrees of freedom which we do control. In such an interpretation, the generalized $\delta Q_k$ is the part of the change of the corresponding expectation value $\delta \langle f_k(x) \rangle$ related to the change in the probability distribution by equation (\ref{eq_Q_k_d_expt_f_k_expt_d_f_k}).  Equations (\ref{eq_Q_k_d_expt_f_k_expt_d_f_k}) and (\ref{eq_dS_sum_k_L_k_Q_k}) explictily show that the change in the maximum of information entropy comes from the change in the probability distribution related to $\delta Q_k$. Furthermore, Grandy has brought generalized terms $\delta Q_k$ into connection with the change of the macroscopic constraints brought by means of the external influences on the system. Based on that, Grandy \cite{grandy,grandy1,grandy2,grandy3} has developed a generalized approach which, along with the generalization of the Liouville--von Neumann equation for the density matrix through the application of the MaxEnt formalism, leads to  the derivation of the macroscopic equations of motion.

Let us consider now the quasistatic change of the energy of macroscopic system, for which we specify only that it is a closed system (i.e. the system that can exchange energy, in the form of work or heat, with the environment, but not particles). From equations  (\ref{eq_dS_sum_k_L_k_Q_k}) and (\ref{eq_F_L_T_dQ_eq_dU_dW}) then it follows that
\begin{equation}
\delta  S = \lambda \delta Q \ ,\label{eq_dS_eq_L_dQ_q_s_e}
\end{equation}   
and
\begin{equation}
\delta U - \delta W = \delta Q  = \frac{1}{\lambda }\delta  S \ . \label{eq_FLT_dU_dW_eq_dQ_L_dS}
\end{equation}  
If we write the first law of thermodynamics in the form in which the thermodynamic entropy $S_e$ explicitly appears,
\begin{equation}
dU - \delta W = \delta Q = TdS_e \ ,
\end{equation} 
then the Lagrange multiplier $\lambda $ in the analogous equation (\ref{eq_FLT_dU_dW_eq_dQ_L_dS}) can be identified as
\begin{equation}
\lambda = \frac{1}{kT} \ . \label{eq_ident_L_eq_kT-1}
\end{equation}
The change $\delta  S$ in the maximum of information entropy given by (\ref{eq_dS_eq_L_dQ_q_s_e}) is thus related to the total differential of thermodynamic entropy $dS_e$ by
\begin{equation}
k d S = d S_e = {\delta Q \over T} \ , \label{eq_kdS_eq_dS_e_eq_Q_T-1}
\end{equation} 
where $T$ is the temperature, and $1/T$ is the integrating factor for heat $\delta Q$. The choice of the unit for temperature (Kelvin), and respectively for entropy (Joule Kelvin$^{-1}$) is reflected in the appearance of the Boltzmann constant $k$ in the previous expressions. 

The confirmation that the identification given by (\ref{eq_ident_L_eq_kT-1}) is correct comes by introducing the value of the Lagrange multiplier $\lambda = (kT)^{-1}$ in the MaxEnt probability distribution corresponding to the case considered here. In this way we obtain 
\begin{equation}
p_i =\frac{1}{Z} \exp \left (- \frac{E_i}{kT}\right ) \ , \label{eq_Gibbs_kan_E_i}
\end{equation}
which is known in statistical mechanics as the Gibbs canonical distribution, describing the closed system of known temperature in equilibrium with the environment. The normalization factor of the canonical distribution, the partition function $Z$, is equal to
\begin{equation}
Z = \sum_{i=1}^{n} \exp \left (- \lambda E_i\right ) = \sum_{i=1}^{n} \exp \left (- \frac{E_i}{kT}\right ) \ . \label{eq_part_fun_Z_L_Z_kT_-1}
\end{equation} 
By considering the open system (i.e. the system that can exchange energy and particles with the environment) in analogous way it is shown that the MaxEnt probability distribution, in the case when along with the expected value of energy, the expected value of the number of particles is known, corresponds to the Gibbs grand canonical distribution \cite{jaynes6,grandy}. Furthermore, it is important that the generic MaxEnt relations from the previous and this section become, in the special cases considered here, the well known equations of equilibrium statistical mechanics.

However, recent work \cite{hanggi} on the Crooks fluctuation theorem \cite{crooks} and Jarzynski equality \cite{jarzynski} indicates further insights. When these important relations of nonequilibrium statistical mechanics are extended to quantum systems strongly coupled with their environments, the thermodynamic entropy of the system of interest in such cases is related to the maximum of the information entropy of the total system (including the system of interest and its environment) minus the information entropy of the environment:
\begin{equation}
S_{e \ \mathrm{of\ the\ system}} = k\left (S_{I\ \mathrm{of\ the\ total\ system}}- S_{I\ \mathrm{of\ the \ environment}}\right )_{\mathrm{max}},
\end{equation}
where $k$ is the Boltzmann constant. The reason for this is that, unlike in the cases considered in this paper, for systems strongly interacting with their environments the correlation between the system and the environment degrees of freedom can not be neglected.

\section{MaxEnt and the interpretation of probabilities} \label{sec_4}

In this section we modify and extend the analysis given by Jaynes in \cite{jaynes8} and show how this leads to the independent interpretation of probabilities which is based on the maximum information entropy principle. Let's consider a proposition $A(n_1, \dots, n_m)$ which is a function of the sample numbers $n_i$, $i = 1, 2, \dots, m$. In the context of statistical mechanics, the sample numbers can represent, for example, the distribution $\{n_1, \dots, n_m\}$  of the number of systems from the ensemble of $n = \sum_{i=1}^m n_i$ identical systems found in $m$ different microscopic states comprising the discrete sample space. The proposition $A(n_1, \dots, n_m)$ can represent, for example, the expected value of energy of the individual system, or the expected values of some set of compatible quantities. Relative frequencies are then given by $f_i = n_i/n$, $i = 1, 2, \dots, m$. The number of outcomes for which the proposition $A$ is true is given by the sum over different distributions of sample numbers $\{n_1, \dots, n_m\}$,
\begin{equation}
M(n, A) = \sum_{\{n_i\} \in R} W(n_1, \dots, n_m) \ , \label{eq_M_sum_W}
\end{equation}
where $R$ is the region of the sample space for which the proposition $A$ is true and $W$ is the multinomial coefficient
\begin{equation}
W(n_1, \dots, n_m) = \frac{n!}{n_1 ! \cdots n_m !} \ .
\end{equation}
The greatest term (multiplicity) in the sum (\ref{eq_M_sum_W}) over the region $R$ is 
\begin{equation}
W_{\mathrm{max}} = \mathrm{Max}_{R}W(n_1, \dots, n_m) \ .  
\end{equation}
If $T(n, m)$ is the number of terms in the sum (\ref{eq_M_sum_W}), then it is true that
\begin{equation}
W_{\mathrm{max}} \leq M(n, A) \leq W_{\mathrm{max}}T(n, m) \ ,
\end{equation}
and
\begin{eqnarray}
&& \frac{1}{n} \log W_{\mathrm{max}} \cr\nonumber\\
&& \leq \frac{1}{n} \log M(n, A) \cr\nonumber\\
&& \leq \frac{1}{n} \log W_{\mathrm{max}} + \frac{1}{n} \log T(n, m) \ . \label{eq_logM_ineq}
\end{eqnarray}
From combinatorial arguments it follows that
\begin{equation}
T(n, m) = {n + m - 1 \choose n} = \frac{(n + m - 1)! }{n! (m - 1)!} \ .
\end{equation}
Then as $n \rightarrow \infty $ 
\begin{equation}
T(n, m) \sim \frac{n^{m - 1}}{(m - 1)!} \ .
\end{equation}
Therefore, as $n \rightarrow \infty $, $\log T(n, m)$ grows less rapidly than $n$, 
\begin{equation}
\frac{1}{n} \log T(n, m) \rightarrow 0 \ , \label{eq_lim_1/n_logT}
\end{equation}
and from (\ref{eq_logM_ineq}) and (\ref{eq_lim_1/n_logT}) it follows that 
\begin{equation}
\frac{1}{n} \log M(n, A) \rightarrow \frac{1}{n} \log W_{\mathrm{max}} \ , \label{eq_log_M_Wmax}
\end{equation}
as $n \rightarrow \infty $. The multinomial coefficient $W$ grows so rapidly with $n$ that the maximum term $W_{\mathrm{max}}$ dominates, in the sense given by (\ref {eq_log_M_Wmax}), the total multiplicity  $M(n, A)$ given by the sum (\ref{eq_M_sum_W}). 

However, the limit we really want is the one in which the sample frequencies $n_i/n$ tend to certain (but not yet specified) constant values $f_i$ as $n \rightarrow \infty $. Therefore, we want the limit of 
\begin{equation}
\frac{1}{n}\log W = \frac{1}{n}\log \left [\frac{n!}{(nf_1) ! \cdots (nf_m) !} \right ] \ ,
\end{equation}
as $n \rightarrow \infty $. Using the Stirling asymptotic approximation
\begin{equation}
\log{n!} \sim n \log n - n + \log \sqrt{2\pi n} + O\left (\frac{1}{n} \right ) \ ,
\end{equation}
we find as $n \rightarrow \infty $ that in this limit we have 
\begin{equation}
\frac{1}{n} \log W \rightarrow H \equiv - \sum_{i=1}^{m} f_i \log f_i \ , \label{eq_lim_logW_H}
\end{equation}
and this gives the information entropy of the relative frequency distribution $\{f_1, \dots, f_m \}$. So,  from (\ref{eq_log_M_Wmax}) and (\ref{eq_lim_logW_H}), it follows that in a such limit we also have that
\begin{equation}
\frac{1}{n} \log M(n, A) \rightarrow \frac{1}{n} \log W_{\mathrm{max}} = H_{\mathrm{max}} \ . \label{eq_log_M_Wmax_H}
\end{equation}
Therefore, for very large $n$, the maximum multiplicity $W_{\mathrm{max}}$ is the one that dominates the total multiplicity $M(n, A)$ and maximizes the information entropy $H$ subject to the constraints that define the region of the sample space for which the proposition $A$ is true. Furthermore, it is straightforward to show that the probability of obtaining the relative frequency distribution $\{f_1, \dots, f_m \}$ which corresponds to the maximum multiplicity $W_{\mathrm{max}}$ approaches $1$ in the limit of large $n$, because from (\ref{eq_log_M_Wmax}) in this limit we have
\begin{equation}
 {W_{\mathrm{max}} \over M(n,A)} \sim 1 \ .
\end{equation} 
Therefore, in the limit of large $n$, without any other additional constraints except that the proposition $A$ is true, we can assume  with certainty that the relative frequencies   
to be used are the ones that maximize the multiplicity $W$ and, because of (\ref{eq_lim_logW_H}) and (\ref{eq_log_M_Wmax_H}), maximize the information entropy $H$. Therefore, according to the weak law of large numbers, the relative frequencies in the limit of a large number of trials ($n \rightarrow  \infty $) should correspond to the MaxEnt probabilities.

So, in this context, can we now examine the frequency interpretation of probabilities as factual properties of the real world, if, as in this example, the corresponding probabilites (obtained in the limit of a large number of trials) actually follow from the principle of maximum information entropy, and therefore, depending only on the available information (i.e. on the proposition $A$), depend on our state of knowledge? This question comes naturally as the above result implies that under constraints representing the information that is available, the relative frequencies in the limit of a large number of trials tend with certainty to the corresponding MaxEnt probabilities.

\section{Conclusion} \label{sec_5}
We have shown how the probabilities in statistical mechanics can not be simply interpreted in the frequentist context. Probabilities, at least in the Gibbs formalism of statistical mechanics, are not simply relative frequencies in the ensemble of a large number of identical systems. Actually they depend on the available information about the individual system and therefore are the description of a degree our knowledge about it. The ensembles of identically prepared systems are chosen in the Gibbs formalism only to illustrate that the information we have about the individual system is incomplete, which means that it isn't sufficiently detailed to specify the exact microscopic state of a macroscopic system, nor its exact evolution in time.   

Furthermore, in the case of nonequilibrium systems and processes that are irreversible on the macroscopic level, justification of nonequilibrium ensembles in the frequentist sense as a physical fact, using only first principles, via equations of motion and ergodic theorems, becomes permeated with technical and, more importantly, conceptual difficulties \cite{jaynes9}. For example, the applications of ergodic theorems for that purpose would require an infinite or large time intervals, and this is not in general always available for nonequilibrium systems that are continuously evolving and changing its macroscopic state with time. This is well exemplified in the work of Zubarev and his coworkers, who introduce a hierarchy of time scales, with different sets of quantities that are relevant for the description of a nonequilibrium system on different time scales \cite{zubarev1,zubarev2}. More important than ergodicity is the concept of a mixing system, originally introduced by Gibbs \cite{gibbs}. Mixing implies ergodicity, and hopefully can provide a mechanical foundation of both nonequilibrium and equilibrium statistical mechanics, if we can prove it for realistic systems \cite{dorfman}. However, there are differing opinions about its importance since transport coefficients and dissipativity, an essential property of macroscopic systems, can not be derived only from mixing \cite{balescu}. On the other hand, as we have shown here, MaxEnt formalism is an independent logical extension of the Gibbs method, and leads to statistical distributions which depend only on the available information. If that information is relevant for the description of a system at a macroscopic level, then accordingly, the obtained statistical distributions should be relevant for describing its macroscopic state, its properties and time evolution \cite{jaynes2,jaynes4,jaynes6,jaynes7,grandy,grandy1,grandy2,grandy3,zubarev1,zubarev2,kuic,kuic1,kuic2,kuic3}.

\end{document}